\documentclass{appolb}
\usepackage{graphicx}

\begin{document}
\title{Elements of a non-Hermitian quantum theory without Hermitian conjugation --- scalar product and scattering%
\thanks{Parts of the comprehensive presentation ``Strong interactions with quarks and mesons --- on unitarisation including bound states and non-Hermitian quantum theory" at the  workshop on ``Unquenched Hadron Spectroscopy:
     Non-Perturbative Models and Methods of QCD vs. Experiment (EEF70)"  (1-5 September, 2014, University of Coimbra, Coimbra, Portugal) ({\sf http://cfif.ist.utl.pt/$\sim$rupp/EEF70/talks/Kleefeld.pdf}).}%
}
\author{Frieder Kleefeld
\address{ Collaborator of the Centro de F\'{\i}sica das Interac\c{c}\~{o}es Fundamentais (CFIF), \\
             Instituto Superior T\'{e}cnico (IST), \\
             Edif\'{\i}cio Ci\^{e}ncia, Piso 3,  
             Av. Rovisco Pais, 
             P-1049-001 LISBOA,
             Portugal\\
	 Private permanent address:\\
 Frankenstr. 3, 91452 Wilhermsdorf, Germany \\
              e-Mail: reste.kleefeld@t-online.de, kleefeld@cfif.ist.utl.pt}
}
\maketitle
\begin{abstract}
The descripition of in a Hermitian setting seemingly nonlocal and nonperturbative phenomena like confinement or superconductivity is most conveniently performed by generalizing quantum theory to a non-Hermitian regime where these phenomena appear perturbative and local. The short presentation provides a clue how this can be done on the basis of Lorentz covariance while preserving the analyticity of the theory. After deriving with the help of Lorentz covariance a quantum scalar product without making any use of metric or complex conjugation we sketch how the formalism of scattering theory can be extended analytically to a non-Hermitian regime.
\end{abstract}
\PACS{03.65.Ca, 03.65.Nk, 03.65.-wi, 11.30.-j, 11.55.-m, 11.80.Gw}
  
\section{Introductory remarks}
To us there are mainly three points becoming gradually clear after about 20 years of intensive reseach effort to generalize quantum theory (QT) to a non-Hermitian (NH) setting (see e.g.\ Refs.\ \cite{Kleefeld:2012}-\cite{Kleefeld:1998yj}
 and references therein): 1) The description of physical systems within an idealized Hermitian setting is at odds with experimental reality; 2) Various strong statements\footnote{E.g.\ that confinement cannot be generated by scalar bosons or that the quartic coupling of a Higgs scalar has --- due to stability reasons --- to be positive \cite{Kleefeld:2005hf}.}  made naively within a Hermitian setting do not hold in a NH setting; 3) The advanced sector of NHQT required by Lorentz covariance \cite{Kleefeld:2012fp} and analyticity is in a Hermitian setting obtained by applying to the retarded sector a Hermitian conjugation joint with a non-local, non-analytic metric \cite{Bender:2002vv}.

\section{Setup of non-Hermitian Quantum Theory (NHQT)}
\subsection{Covariance and conservation of complex energy in the complex plane}
In the first place one should recall that the relativistic Klein-Gordon equation being essentially the wave equation is a differential equation of second order in the time coordinate which can be decomposed  \cite{Kleefeld:2012fp,Kleefeld:2006bp,Kleefeld:2004qs} into two first order equations. For a --- without loss of generality --- time independent eventually NH Hamilton operator $H$ we have:
\begin{equation}  0 =  \left(( i\,\hbar)^2 \, \frac{\partial^2}{\partial t^2}   - H^2 \; \right)   \big|\psi (t) \big>   =  \left( i\,\hbar \, \frac{\partial}{\partial t}   - H \; \right)  \left( i\,\hbar \, \frac{\partial}{\partial t}   + H \; \right)  \big|\psi (t) \big> \; .   
\end{equation}
The right solution of the Klein-Gordon equation $\big|\psi (t) \big>=\big|\psi^{\,(+)} (t) \big>+\big|\psi^{\,(-)} (t) \big>$ is therefore obtained by superimposing additively the solutions $\big|\psi^{(+)} (t)\big>$ and $\big|\psi^{(-)} (t)\big>$ of the retarded and advanced Schr\"odinger equation respectively:
\begin{equation}
  0 =  \left( i\,\hbar \, \frac{\partial}{\partial t}   -  H \; \right) \; \big|\psi^{(+)} (t)\big> \, , \quad  0  =  \left( i\,\hbar \, \frac{\partial}{\partial t}   + H \; \right)\; \big|\psi^{(-)} (t)\big>   \, . \label{sequ1} \end{equation}
The respective left eigen-solution $\big<\!\big<\psi^{(+)} (t)\big|$ and $\big<\!\big<\psi^{(-)} (t)\big|$  of these two equations is the right eigen-solution $\big|\psi^{(+)} (t)\big)\equiv\big<\!\big<\psi^{(+)} (t)\big|^T$ and $\big|\psi^{(-)} (t)\big)\equiv\big<\!\big<\psi^{(-)} (t)\big|^T$ of the respective so-called ``transposed retarded" and ``transposed advanced" Schr\"odinger equation (Here $T$ denotes transpositon!), i.e.:
\begin{equation}
  0 =  \left( i\,\hbar \, \frac{\partial}{\partial t}   -  H^T \; \right) \; \big|\psi^{(+)} (t)\big) \, , \quad  0  =  \left( i\,\hbar \, \frac{\partial}{\partial t}   + H^T \; \right)\; \big|\psi^{(-)} (t)\big)   \, . \label{tsequ1} \end{equation}
In using the notation $\psi^{(\pm)}_R(z,t)\equiv\big<\!\big<z\big|\psi^{(\pm)} (t)\big>\equiv \big(\!\big(\psi^{(\pm)} (t)\big|z\big)^T$ and $\psi^{(\pm)}_L(z,t)\equiv\big(\!\big(z\big|\psi^{(\pm)} (t)\big)\equiv \big<\!\big<\psi^{(\pm)} (t)\big|z\big>^T$ the non-relativistic one-dimensional limit of Eqs.\ (\ref{sequ1}) and (\ref{tsequ1}) reads in spatial representation:
\begin{eqnarray} 
 \pm\, i\,\hbar \, \frac{\partial}{\partial t}  \, \psi_R^{(\pm)}(z,t)  & = &   \left( - \frac{\hbar^2}{2\,M} \frac{\partial^2}{\partial z^2} + V(z) \right)\psi_R^{(\pm)}(z,t) \, , \label{sequ2} \\
 \pm\, i\,\hbar \, \frac{\partial}{\partial t}  \, \psi_L^{(\pm)}(z,t)  & = &   \left( - \frac{\hbar^2}{2\,M} \frac{\partial^2}{\partial z^2} + V(z)^T \right)\psi_L^{(\pm)}(z,t) \, .  \label{sequ3}
\end{eqnarray}
On the basis of these equations it is easy to show that there hold the following two continuity equations
\begin{equation} \frac{\partial\, \rho^{(+)}(z,t)}{\partial t}  \; = \; - \, \frac{\partial j^{(+)}(z,t)}{\partial z}\; , \quad  \frac{\partial\, \rho^{(-)}(z,t)}{\partial t}  \; = \; - \, \frac{\partial j^{(-)}(z,t)}{\partial z} \label{conteq1}   
\end{equation}
for the retarded and advanced energy densities $\rho^{(+)}(z,t)$ and $\rho^{(-)}(z,t)$ and the respective energy current densities $j^{(\pm)}(z,t)$ defined as follows:
\begin{eqnarray} \lefteqn{\rho^{(\pm)}(z,t) \;  = \;  \psi_L^{(\mp)}(z,t)^T\! \cdot \psi_R^{(\pm)}(z,t) \; ,}  \\[2mm] 
\lefteqn{j^{(\pm)}(z,t) \; =} \nonumber \\[2mm]
  & = & \frac{1}{\pm\, i\hbar} \left( \psi_L^{(\mp )}(z,t)^T\! \cdot \frac{\hbar^2}{2\,M} \frac{\partial  \psi_R^{(\pm )}(z,t)}{\partial z} -  \frac{\partial \psi_L^{(\mp )}(z,t)^T}{\partial z}   \cdot \frac{\hbar^2}{2\,M}\,\psi_R^{(\pm )}(z,t) \right)  . \nonumber \\
\label{currdens1} \end{eqnarray}
The continuity equations (\ref{conteq1}) can be integrated along some suitable contour connecting two points $z_1$ and $z_2$ in the complex $z$-plane yielding
\begin{equation} \frac{\partial}{\partial t} \int_{z_1}^{z_2} dz \; \rho^{(\pm)}(z,t)  =  - \, \Big( j^{(\pm)}(z_2,t) -  j^{(\pm)}(z_1,t)  \Big) \; . 
\end{equation}
Any integration contour with $j^{(\pm)}(z_2,t) =  j^{(\pm)}(z_1,t)$ defines an eventually NHQT with a time-independent scalar product \cite{Kleefeld:2012,Kleefeld:2004qs}
\begin{equation}  \int_{z_1}^{z_2} dz \; \rho^{(\pm)}(z,t)  =   \int_{z_1}^{z_2} dz \;     \psi_L^{(\mp)}(z,t)^T\! \cdot \psi_R^{(\pm)}(z,t) = \mbox{const} 
\end{equation}
replacing the well known scalar product of Max Born.

\subsection{Elements of non-Hermitian scattering theory}
Without loss of generality we consider now one-dimensional scattering at a time-independent eventually NH potential $V(z)$. For such a potential the Schr\"odinger equations (\ref{sequ2}) and (\ref{sequ3}) can be solved by a separation ansatz $\psi_R^{(\pm )}(z,t) = \exp (\pm E t/(i\hbar)) \,\phi_R^{(\pm )}(z)$ and  $\psi_L^{(\pm )}(z,t) = \exp (\pm E t/(i\hbar)) \,\phi_L^{(\pm )}(z)$ yielding the time-independent Schr\"odinger equations
\begin{eqnarray} 
 E  \; \phi_R^{(\pm)}(z)  & = &   \left( - \frac{\hbar^2}{2\,M} \frac{\partial^2}{\partial z^2} + V(z) \right)\phi_R^{(\pm)}(z) \, , \label{sequ4} \\
 E  \; \phi_L^{(\pm)}(z)  & = &   \left( - \frac{\hbar^2}{2\,M} \frac{\partial^2}{\partial z^2} + V(z)^T \right)\phi_L^{(\pm)}(z) \, ,  \label{sequ5}
\end{eqnarray}
and according to Eqs.\ (\ref{currdens1}) the time-independent energy current densities
\begin{equation} 
j^{(\pm)}(z)  = \frac{1}{\pm\, i\hbar} \left( \phi_L^{(\mp )}(z)^T\! \cdot \frac{\hbar^2}{2\,M} \frac{\partial  \phi_R^{(\pm )}(z)}{\partial z} -  \frac{\partial \phi_L^{(\mp )}(z)^T}{\partial z}   \cdot \frac{\hbar^2}{2\,M}\,\phi_R^{(\pm )}(z) \right)  . 
\end{equation}
Here we will discuss merely retarded scattering. Hence we use in the following the abbreviations $j(z)\equiv j^{(+)}(z)$, $\phi^{(+)}(z)\equiv\phi^{(+)}_R(z)$, $\phi^{(-)}(z)\equiv\phi^{(-)}_L(z)$,  $\phi^{(+)}(z)^\prime \equiv\partial \phi^{(+)}_R(z)/\partial z$ and $\phi^{(-)}(z)^\prime \equiv\partial \phi^{(-)}_L(z)/\partial z$. Advanced scattering results are nonetheless easily derivable from their retarded counterparts. In the region of vanishing potential, i.e. $V(z)=0$, the solution of the Schr\"odinger Eqs. (\ref{sequ4}) and (\ref{sequ5}) is of plane-wave type with $k_0\equiv \sqrt{2ME/\hbar^2}$:
\begin{eqnarray}  \phi^{(\pm)}(z) & = & \exp(\pm i\, k_0z)\; c^{(\pm)}(k_0)+\exp(\mp i\, k_0z)\: c^{(\pm)}(-k_0) \, , \\
\Rightarrow \; j(z) & = & \frac{1}{ i\hbar} \Big( \phi^{(-)}(z)^T\! \cdot \frac{\hbar^2}{2\,M} \; \phi^{(+)}(z)^\prime -  \phi^{(-)}(z)^{\prime \, T}   \cdot \frac{\hbar^2}{2\,M}\;\phi^{(+)}(z) \Big) \nonumber \\
 & = &  c^{(-)}(k_0)^T\! \cdot \frac{\hbar k_0}{M} \; c^{(+)}(k_0) -  c^{(-)}(-k_0)^T\! \cdot \frac{\hbar k_0}{M} \; c^{(+)}(-k_0) \; . \quad \label{currdens2} \end{eqnarray}
In the following we want to consider retarded scattering between two points $z_<$ and $z_>$ of vanishing potential, i.e.\ $V(z_<)=V(z_>)=0$. Wave functions and their derivatives at $z_>$ and  $z_<$ are related by transfer matrices $T^{(\pm)}$:
\begin{equation} \left( \begin{array}{l} \sqrt{\frac{\hbar^2}{2M}}\; \phi^{(\pm)}(z_>) \\[1mm]  \sqrt{\frac{\hbar^2}{2M}}\; \phi^{(\pm)}(z_>)^\prime \end{array}\right)
  =   \left( \begin{array}{cc} T^{(\pm)}_{11} & T^{(\pm)}_{12} \\[1mm]  T^{(\pm)}_{21} & T^{(\pm)}_{22} \end{array} \right) \, \left( \begin{array}{l} \sqrt{\frac{\hbar^2}{2M}}\; \phi^{(\pm)}(z_<) \\[1mm]  \sqrt{\frac{\hbar^2}{2M}}\; \phi^{(\pm)}(z_<)^\prime \end{array}\right)  ,
\end{equation}
or, alternatively,
\begin{equation} \left( \begin{array}{l}  a^{(\pm)}_1 \\[1mm] e^{(\pm)}_2  \end{array}\right)
  =  \tilde{T}^{(\pm)} \left( \begin{array}{l}  e^{(\pm)}_1 \\[1mm] a^{(\pm)}_2  \end{array}\right)   =   \left( \begin{array}{cc} \tilde{T}^{(\pm)}_{11} & \tilde{T}^{(\pm)}_{12} \\[1mm]  \tilde{T}^{(\pm)}_{21} & \tilde{T}^{(\pm)}_{22} \end{array} \right) \left( \begin{array}{l}  e^{(\pm)}_1 \\[1mm] a^{(\pm)}_2  \end{array}\right) , \label{ttild1}
\end{equation}
with
\begin{eqnarray} e^{(\pm)}_1  \equiv  e^{\pm ik_0z_<}\; \sqrt{\frac{\hbar k_0}{M}} \; c^{(\pm)}_<(k_0) & , & e^{(\pm)}_2  \equiv  e^{\mp ik_0z_>}\; \sqrt{\frac{\hbar k_0}{M}} \; c^{(\pm)}_>(-k_0) \; , \\
 a^{(\pm)}_1  \equiv  e^{\pm ik_0z_>}\; \sqrt{\frac{\hbar k_0}{M}} \; c^{(\pm)}_>(k_0) & , & a^{(\pm)}_2  \equiv  e^{\mp ik_0z_<}\; \sqrt{\frac{\hbar k_0}{M}} \; c^{(\pm)}_<(-k_0) \; .
\end{eqnarray}
Simple algebra establishes the following relation between $\tilde{T}^{(\pm)}$ and $T^{(\pm)}$:
\begin{equation} \tilde{T}^{(\pm)} = 1_2 + \frac{\sqrt{k_0}}{2} \left( \begin{array}{cc} 1 & \pm \frac{1}{ik_0} \\[1mm]  1 &  \mp \frac{1}{ik_0} \end{array} \right) ( T^{(\pm)} -1_2) \left( \begin{array}{cc} 1 & 1 \\[1mm]  \pm i k_0 & \mp ik_0 \end{array} \right) \frac{1}{\sqrt{k_0}} \; . \label{ttrel1}
\end{equation}
$1_2$ is the $2\times 2$ unit matrix.
Moreover we assume the energy current densities at  points $z_<$ and $z_>$ to be equal, i.e.\ $j(z_<)=j(z_>)$, yielding (see Eq. (\ref{currdens2}))
\begin{eqnarray}
\lefteqn{ c^{(-)}_>(k_0)^T\! \cdot \frac{\hbar k_0}{M} \; c^{(+)}_>(k_0) -  c^{(-)}_>(-k_0)^T\! \cdot \frac{\hbar k_0}{M} \; c^{(+)}_>(-k_0)=} \nonumber \\
 & = &  c^{(-)}_<(k_0)^T\! \cdot \frac{\hbar k_0}{M} \; c^{(+)}_<(k_0) -  c^{(-)}_<(-k_0)^T\! \cdot \frac{\hbar k_0}{M} \; c^{(+)}_<(-k_0) \; , \end{eqnarray}
or, equivalently, $a^{(-)\,T}_1\! \cdot \, a^{(+)}_1 -  e^{(-)\,T}_2\! \cdot \, e^{(+)}_2 =  e^{(-)\,T}_1\! \cdot \, e^{(+)}_1 -  a^{(-)\,T}_2\! \cdot \, a^{(+)}_2$. 
Inspecting $e^{(-)\,T}_1\! \cdot \, e^{(+)}_1 +  e^{(-)\,T}_2\! \cdot \, e^{(+)}_2 =  a^{(-)\,T}_1\! \cdot \, a^{(+)}_1 +  a^{(-)\,T}_2\! \cdot \, a^{(+)}_2$ we can define the S-matrix $S^{(+)}$ and transpose of its  inverse $S^{(-)}=(S^{(+)-1})^T$ by
\begin{equation} \left( \begin{array}{l}  a^{(\pm)}_1 \\[1mm] a^{(\pm)}_2  \end{array}\right)
  =  S^{(\pm)} \left( \begin{array}{l}  e^{(\pm)}_1 \\[1mm] e^{(\pm)}_2  \end{array}\right)   =   \left( \begin{array}{cc} S^{(\pm)}_{11} & S^{(\pm)}_{12} \\[1mm]  S^{(\pm)}_{21} & S^{(\pm)}_{22} \end{array} \right) \left( \begin{array}{l}  e^{(\pm)}_1 \\[1mm] e^{(\pm)}_2  \end{array}\right) .
\end{equation}
Making use of Eqs.\ (\ref{ttild1}) and $S^{(-)T}S^{(+)}=1_2$ we obtain:
\begin{eqnarray} S^{(\pm)} \! = \! \left( \begin{array}{cc} S^{(\pm)}_{11} & S^{(\pm)}_{12} \\[1mm]  S^{(\pm)}_{21} & S^{(\pm)}_{22} \end{array} \right) & = &  \left( \begin{array}{ll} \left(\tilde{T}^{(\mp)T}_{11}\right)^{-1} & \tilde{T}^{(\pm)}_{12} \tilde{T}^{(\pm)-1}_{22} \\[1mm]  -\tilde{T}^{(\pm)-1}_{22}\tilde{T}^{(\pm)}_{21} & \tilde{T}^{(\pm)-1}_{22} \end{array} \right) . \\
 S^{(\mp)T} \! = \! \left( \begin{array}{cc} S^{(\mp)T}_{11} & S^{(\mp)T}_{21} \\[1mm]  S^{(\mp)T}_{12} & S^{(\mp)T}_{22} \end{array} \right) & = &  \left( \begin{array}{ll} \tilde{T}^{(\pm)-1}_{11}  & -\tilde{T}^{(\pm)-1}_{11} \tilde{T}^{(\pm)}_{12} \\[1mm]  \tilde{T}^{(\pm)}_{21}\tilde{T}^{(\pm)-1}_{11} & \left(\tilde{T}^{(\mp)T}_{22}\right)^{-1} \end{array} \right) .
\end{eqnarray}
The transmittivities $T_1$ and $T_2$ and reflectivities $R_1$ and $R_2$ are therefore:
\begin{eqnarray} T_1 = 1 - R_1 = S^{(-)T}_{11} S^{(+)}_{11} & = & \tilde{T}^{(+)-1}_{11}   \left(\tilde{T}^{(-)T}_{11}\right)^{-1}=  \left(\tilde{T}^{(-)T}_{11} \tilde{T}^{(+)}_{11}  \right)^{-1} , \quad \\
 T_2 = 1 - R_2 = S^{(-)T}_{22} S^{(+)}_{22} & = &  \left(\tilde{T}^{(-)T}_{22}\right)^{-1} \tilde{T}^{(+)-1}_{22}  =  \left(\tilde{T}^{(+)}_{22} \tilde{T}^{(-)T}_{22}   \right)^{-1} . \quad 
\end{eqnarray}
\section{Simple application: scattering at a delta-potential}
For the scattering at a delta-potential $V(z)=g\;\delta(z-a)$ with $g$ being eventually complex-valued we choose $x_>=a+0$ and $x_<=a-0$. The delta-potential is represented by the following transfer matrices:
\begin{equation} T^{(+)} = \left( \begin{array}{ll} 1 & 0 \\[1mm]
 \sqrt{\frac{2M}{\hbar^2}} \; g \; \sqrt{\frac{2M}{\hbar^2}} & 1 \end{array} \right)\; , \;  T^{(-)} = \left( \begin{array}{ll} 1 & 0 \\[1mm]
 \sqrt{\frac{2M}{\hbar^2}} \; g^T \; \sqrt{\frac{2M}{\hbar^2}} & 1 \end{array} \right) \; .
\end{equation}
Invoking these transfer matrices into Eq.\ (\ref{ttrel1}) we obtain
\begin{eqnarray} \tilde{T}^{(+)}_{11} = \tilde{T}^{(-)T}_{22} = & \left(S^{(-)T}_{11}\right)^{-1} = \left(S^{(-)T}_{22}\right)^{-1}  & =  1 + \frac{1}{2i}\; \sqrt{\frac{2M}{\hbar^2 k_0}} \; g \; \sqrt{\frac{2M}{\hbar^2k_0}}\,, \; \quad \\
 \tilde{T}^{(-)T}_{11} = \tilde{T}^{(+)}_{22} = & \left(S^{(+)}_{11}\right)^{-1} = \left(S^{(+)}_{22}\right)^{-1}  & =  1 - \frac{1}{2i}\; \sqrt{\frac{2M}{\hbar^2 k_0}} \; g \; \sqrt{\frac{2M}{\hbar^2k_0}}\,, \; \quad 
\end{eqnarray}
and, consequently, 
\begin{eqnarray} \lefteqn{T_1 = T_2 = 1-R_1 = 1- R_2 =} \nonumber \\
 & = & \left[ \left( 1 - \frac{1}{2i}\; \sqrt{\frac{2M}{\hbar^2 k_0}} \; g \; \sqrt{\frac{2M}{\hbar^2k_0}}\right)\left( 1 + \frac{1}{2i}\; \sqrt{\frac{2M}{\hbar^2 k_0}} \; g \; \sqrt{\frac{2M}{\hbar^2k_0}}\right) \right]^{-1}  .
\end{eqnarray}
This is --- without involving any complex conjugation --- the standard result which will be for one scattering channel obviously real-valued, positive and within the invervall $[0,1]$, if  $(Mg/k_0)^2$ is real-valued and non-negative.

\noindent {\bf Acknowledgements}

Best wishes to Everadus Johannes H.V.\  van Beveren (``Eef") on the occasion of his 70th birthday. Cordial thanks to him and his family for lasting support. We also would like to deliver our best regards for kindest hospitality to  Maria do C\'{e}u Martins Alfaiate Reste in the Rua da Bica 10, 3060-295 Porto dos Cov\~{o}es, Portugal, where essential parts of the presented work have been completed.
This work was supported by the Funda\c{c}\~{a}o para a Ci\^{e}ncia e a Tecnologia of the 
Minist\'{e}rio da Ci\^{e}ncia, Tecnologia e Ensino Superior of Portugal, under contract
CERN/FP/123576/2011  and by the Czech project
LC06002.

\end{document}